\documentstyle[twocolumn,aps]{revtex}

\begin{document}
\title
{Collective Modes in Strongly Coupled Elecronic Bilayer Liquids }
\author{
G. Kalman($^{\star}$), V. Valtchinov($^{\dag,\star}$), and K. I. Golden($^{\ddag}$)}
\address{\\
($^\star $) Department of Physics, Boston College, Chestnut Hill, MA 02467\\
($^\dag $) Department of Radiology, Brigham and Women's Hospital,\\
Harvard Medical School, Boston, MA 02115\\
($^\ddag $) Department of Mathematics and Statistics, University of Vermont,\\
Burlington, VT 05401}
\date{Received \today}


\begin{abstract}
We present the first reliable calculation of the collective mode structure
of a strongly coupled electronic bilayer. The calculation is based on a
classical model through the $3^{rd}$ frequency-moment-sum-rule preserving
Quasi Localized Charge Approximation, using the recently calculated
Hypernetted Chain pair correlation functions. The spectrum shows an energy
gap at $k=0$ and the absence of a previously conjectured dynamical
instability.
\end{abstract}


Electronic bilayers exhibit a rich pattern of behavior, both on the static
and on the dynamic level. While at high $r_{s}$ values ($%
r_{s}>r_{s}^{crystal}$), the bilayer is expected to crystallize (according
to \cite{1,2}, $r_{s}^{crystal}\geq 20$), and at very low $r_{s}$ values the
RPA description is largely sufficient, the most interesting behavior occurs
in the $1<r_{s}<r_{s}^{crystal}$ domain, where the system is in the liquid
state. This is the domain we focus on in this Letter.

The most remarkable feature on the static level is that the system exhibits
a series of abrupt structural changes as the ratio of the interlayer
distance $d$ and the 2D Wigner-Seitz radius $a$ is varied \cite{3}. These
structural changes parallel the structural phase transitions in the solid
phase \cite{4,2}, but at finite temperature they are also combined with
entropy increasing substitutional order-disorder transitions \cite{3} where
particles in layer $1 (2)$ occupy positions appropriate for particles in
layer $2 (1)$: this is signalled by the two pair correlation functions (PCF) 
$h_{11}(r)$ and $h_{12}(r)$ becoming identical as $d\rightarrow 0$.

The dynamical behavior, the collective mode structure in particular, has
been studied by Swierkowski {\it et al.} \cite{5}, by Gold \cite{6}, by
Zhang and Tzoar \cite{7}, Golden, Kalman and collaborators \cite{8,9,10} and
by Moudgil {\it et al.} \cite{11}. (Some of these studies pertain to a
superlattice [infinite number of layers], rather than to a bilayer; in
qualitative terms the results can, however, be easily interpreted for the
bilayer). Classical bilayer and multilayer structures that form in charged
particle traps have also been studied theoretically by Dubin [12a] and have
recently been observed in ion traps [12b]. There are two problematic issues
that make the predictions based on the calculations less than reliable \cite
{13}. The first issue relates to the approximation technique used: most of
the works cited \cite{5,6,7,11} use methods which violate the $3^{rd}$
frequency-moment-sum-rule, whose satisfaction is well recognized \cite{14}
to be an important criterion for providing an acceptable description of the
collective mode behavior. The second issue concerns the use of the
intralayer and interlayer correlation functions as inputs in all the
calculations cited. No reliable pair correlation function (PCF) data -
either for classical or quantum bilayer systems - have been available until
fairly recently: thus predictions of the collective mode structure (which
turns out to be extremely sensitive to the behavior of the inputted PCF)
have been compromised from the outset.

In this Letter we present the first consistent and reliable calculation of
the collective mode spectrum of a strongly coupled electronic bilayer
liquid. Our results show features which are qualitatively different from the
weakly coupled RPA results; they also show that earlier claims concerning
the possible emergence of a dynamical instability \cite{5,6,9,11} cannot be
supported by a more consistent treatment of the correlations. The
calculation is based on a purely classical model: (i) two 2D electron
liquids separated by distance $d$; (ii) scattering on impurities etc.
neglected; (iii) no interlayer tunneling; (iv) the system is described as a
binary liquid with interaction potentials $\varphi_{11}(r)=%
\varphi_{22}(r)=e^2/r, \varphi_{12}(r)=e^{2}/ \sqrt{(r^2+d^2)}$. In
addition, the model portrays a classical electron liquid where exchange and
other quantum effects are neglected: this approximation is reasonable in the
strong coupling domain where the particles are well localized. The
intralayer coupling is characterized by the parameter $\Gamma=e^2/(aT)$
where $a=1/ \sqrt(\pi n)$ and $T$ is the kinetic energy per particle -- the
temperature in a classical system and $(1/2)\varepsilon_{F}$ in a zero
temperature 2D electron gas. Hence the equivalence $\Gamma \rightarrow 2r_s$%
. The calculation of the dielectric matrix ${\bf \varepsilon}_{ij}^{\mu \nu}
({\bf k} \omega)$ is carried out in the Quasi Localized Charge Approximation
(QLCA), which has been applied successfully for the description of other
strongly coupled Coulomb systems \cite{15,16,17}; in the QLCA ${\bf %
\varepsilon}({\bf k} \omega)$ becomes a functional of the intralayer and
interlayer PCF-s $h_{11}(r)$ and $h_{12}(r)$ or of the corresponding
structure functions $S_{11}({\it k})$ and $S_{12}({\it k})$:

\begin{equation}
{\bf \varepsilon}({\bf k} \omega) = {\bf I} - \omega_0^2(ka)[\omega^2{\bf I}
- {\bf D}({\bf k})]^{-1}
\end{equation}

($\omega _{0}=(\frac{2\pi e^{2}n}{ma})^{1/2}$, the nominal plasma frequency
of a single 2D layer) with \cite{15} \FL
\begin{eqnarray}
&&{\bf {\it D}}_{ij}^{\mu \nu }({\bf k})=  \nonumber \\
&&\frac{1}{mA}\sum\limits_{{\bf q}}{q^{\mu }q^{\nu }[\varphi _{ij}({\it q}%
)S_{ij}(|{\bf k-q}|)-\delta _{ij}\sum\limits_{l}\varphi _{il}({\it q})S_{il}(%
{\it q})]},  \label{Ds}
\end{eqnarray}
${\bf I}$ is the identity matrix. In Cartesian space ${\bf \varepsilon }(%
{\bf k}\omega )$ and similarly ${\bf D}({\bf k})$ are reducible to
longitudinal ${\bf \varepsilon }^{L}({\bf D}^{L})$ and transverse ${\bf %
\varepsilon }^{T}({\bf D}^{T})$ matrices. The dispersion relation for the
longitudinal modes is then obtained from

\begin{equation}
||{\bf \varepsilon}^L({\bf k}\omega)|| = 0  \label{epsL}
\end{equation}

which leads to

\begin{equation}
\omega ^{2}=\omega _{0}ka({1\pm e^{-kd})}+{\ D_{11}^{L}({\bf k})\pm
D_{12}^{L}({\bf k})}  \label{omegaL}
\end{equation}

With the neglect of retardation effects, the dispersion relation for the
transverse modes is derived from

\begin{equation}
||{\bf \varepsilon }^{T}({\bf k}\omega )^{-1}||=0  \label{epsT}
\end{equation}
which yields 
\begin{equation}
\omega ^{2}=D_{11}^{T}({\bf k})\pm D_{12}^{T}({\bf k})  \label{omegaT}
\end{equation}
The ${\bf D}$-functions are to be expressed in terms of the structure
functions, according to Eq. \ref{Ds}. These latter have recently been
calculated \cite{3} through the HNC (Hypernetted Chain) integral equation
for a wide range of $\Gamma $ and $d$ values. Inputting them into the
corresponding ${\bf D}$-functions and using the latter in Eqs. \ref{omegaL}
and \ref{omegaT} one can generate a full description of the collective mode
spectrum. The results are portrayed in Figs. 1-3 and the qualitative
features of the collective mode dispersion are summarized below. (The
figures are given for $\Gamma =40$, corresponding to $r_{s}=20$: this $r_{s}$
value, while high enough for correlations to be dominant is within the
domain of experimental realizability.)

1. The spectrum of collective modes comprises 4 modes: 2 (longitudinal and
transverse) in-phase modes (corresponding to the + sign in Eqs. \ref{omegaL}
and \ref{omegaT}) and 2 (longitudinal and transverse) out-of-phase modes
(corresponding to the - sign in Eqs. \ref{omegaL} and \ref{omegaT}).

2. The in-phase modes (where the two layers oscillate in unison) are not
qualitatively different from the similar modes of an isolated 2D layer \cite
{16}. In particular, for $k \rightarrow 0$ the longitudinal (plasmon) mode
has the typical, quasi-acoustic $\omega \sim \sqrt{k}$ dispersion, while the
transverse (shear) mode is acoustic, $\omega \sim k$; both modes are
softened by intralayer and interlayer correlations.

3. The out-of-phase modes (where the oscillations of the two layers exhibit
a $180^{\circ}$ phase difference) are characterized by an $\omega(k=0) > 0$
energy gap. The physical reason for the existence of an energy gap for
layered systems has already been discussed elsewhere \cite{8}. The
out-of-phase longitudinal mode in the RPA approximation has been identified
as the {\it acoustic plasmon} \cite{18,19} since for $k \rightarrow 0$, $%
\omega \rightarrow 0$ as $k$. The present calculation clearly shows the
marked difference brought about by the strong correlations. Since at $k=0$
the isotropy of the system is unbroken, the plasmon and the shear modes
share a common gap value.

4. From Eqs. \ref{omegaL} and \ref{omegaT} the gap value can be expressed as

\begin{equation}
\omega^2(0,d)=-\frac{\omega_0^2}{2} \int\limits_0^{\infty}d(qa)(qa)^2
e^{-qd}S_{12}(q).
\end{equation}

With increasing $d$ and consequently decreasing interlayer correlations, $%
\omega(0)$ shows a decreasing tendency and it virtually vanishes for $d > 1.5
$ when the separated layers become practically uncorrelated \cite{3}. This
downward trend is, however, preceded by a slight upturn for $0<d<0.16$ (for $%
\Gamma \geq 30$). Although the details of this behavior are not well
understood, it is most likely due to the {\it substitutional disorder} that
prevails in this region \cite{3}: the eigenfrequencies of the localized
modes in the substitutionally disordered phase are expected to be higher
than in the substitutionally ordered phase \cite{20}. Within the domain
investigated, the $\Gamma$-dependence of $\omega(0)$ is quite mild, but the
QLCA being a strong coupling approximation, no inference concerning the
behavior of $\omega(0)$ in the moderately coupled ($\Gamma < 10$) domain can
be drawn from this observation. In fact, it is expected that for low enough $%
\Gamma$ values, $\omega(0)$ tends to zero to match at $\Gamma=0$ the
predicted RPA behavior \cite{18}.

5. For finite $k$ values all the four dispersion curves develop an
oscillatory behavior, generated by the similar behavior of the inputted
structure functions. This behavior has also been identified for the isolated
2D layer [16, 17b]. The structure of the out-of-phase plasmon mode is of
special interest here: the first sharp roton-like minimum has attracted
attention in earlier studies \cite{5,9,11} which were based on the neglect
or on a highly approximate treatment of the interlayer correlations. It was
suggested that the minimum of $\omega ^{2}$ may dip below $\omega ^{2}=0$
[5a, 9,11] or may, at least, reach the close vicinity of $\omega ^{2}=0$
[5b,5c]. The former behavior would indicate a dynamical instability
(heralding the onset of CDW-type ground state), [5a, 11]; the latter has
been interpreted as the onset of a new high-$k$, low frequency mode [5b,
5c]. Our results show that the roton minimum never drops below the value
already reached by the dispersion curve of the 2D layer. The consistent
treatment of the interlayer and intralayer correlations thus precludes the
existence of the effects conjectured in \cite{5,9,11}, virtually
independently of any other approximation used.

6. At high {\it k}-values, for a given $d$ all the dispersion curves
approach the same asymptotic frequency value, the frequency of a localized
mode, a particle oscillating in the screening environment of the two layers:

\begin{eqnarray}
\omega ^{2}(\infty ,d) &=&\frac{-1}{mA}\sum\limits_{{\bf q}}[\varphi
_{11}(q)({\it S}_{11}(q)-1)+\varphi _{12}(q){\it S}_{12}({\it q})]q^{2} 
\nonumber \\
&=&-\frac{\omega _{0}^{2}}{4}\int\limits_{0}^{\infty
}d(qa)(qa)^{2}[S_{11}(q)-1+S_{12}(q)e^{-qd}]
\end{eqnarray}

This result is, probably, only of academic interest, since it is unlikely
that high-{\it k} modes would survive the damping mechanism operating in the
system.

The QLCA is not geared to describe damping processes and therefore our
calculation fails to provide information on the damping of the collective
modes. However, some qualitative statements can be made. We concentrate on
the out-of-phase modes only. There are three major damping mechanisms to be
considered. These are (i) single pair excitation (Landau damping), (ii)
multiple pair excitations and (iii) diffusive-migrational damping \cite{17}.
The effect of the single pair excitations can be easily assessed from Figs.
1 and 2. Both of them show the pair excitation domain: it is clear that as
long as the layer separation is not too large ($d/a<1.5$) for small k values
both the out-of-phase plasmon and shear modes are well outside the continuum
and are thus immune to Landau-damping. In a highly correlated plasma
multi-particle excitations are, however, also operative, with increasing
importance at higher $k$ values. The $k\rightarrow \infty $ portion of the
dispersion curve emerging from the continuum would probably be heavily
damped by this process. The diffusive-migrational shifting of the
quasi-sites \cite{17} can originate from quasi-thermal diffusion or from
tunneling between neighboring minima of the fluctuating potential. For $%
\Gamma $ sufficiently high ($\Gamma \ge 40$, i.e. $r_{s}\ge 20$, probably
close to crystallization \cite{1}), the latter effect should be
significantly diminished. The former, however, is sizable when its
characteristic time becomes comparable with the period of the mode under
consideration. Based on our earlier estimate \cite{10} of the lowest
surviving oscillation frequency, $\omega _{min}$ and on the MD results of
Ref. \cite{21}(see also \cite{17}) concerning the lowest propagating
wave-number value for the shear mode, one can conclude that for $\Gamma =40$
($r_{s}=20$) the $d/a<0.8$ domain and for $\Gamma =20$ ($r_{s}=10$) the $%
d/a<0.6$ domain can be safely assumend not to be seriously affected by this
damping mechanism for either of the modes. (cf Fig. 1/a). For higher $d$
values the gap frequency $\omega (0)$ is below $\omega _{min}$ and thus the
shear mode would be propagating for $\omega >\omega _{min}$ only; in the
domain $\omega <\omega _{min}$ the longitudinal mode would be stripped of
its correlational features and would revert to an RPA type acoustic plasmon
with $\omega (k\rightarrow 0)\rightarrow 0$. (cf. Fig. 1/b).

The collective mode structure of the strongly coupled bilayer liquid
presented in this Letter bears a close relationship to the phonon spectrum
of the bilayer solid, recently calculated by Goldoni and Peeters \cite{4}.
The four modes in the solid phase can be identified as the transverse
acoustic and the longitudinal quasi-acoustic ($\sim \sqrt{k} $) phonons and
the transverse and longitudinal optical phonons. The ``gaps`` exhibited by
the latter are, in general, different because of the anisotropy of the
lattice. In contrast, in the liquid state, there is only one single
isotropic gap, as determined in this Letter. This liquid gap value is
typically slightly above the arithmetic average of the optical frequencies
of the transverse and longitudinal phonons, in the solid phase.

Concerning the possible observation of the features predicted in this
Letter, one can suggest three main areas that should lend themselves to
direct experimental verification: (1) the existence and the nonmonotonic $d$%
-dependence of the $k=0$ energy gap; (2) the existence of a transverse shear
excitation with a high frequency and expected low damping (this is in a
sharp contrast to the usual scenario for the shear mode in the liquid phase,
which vanishes for $k \rightarrow 0$ \cite{17,21,22}; (3) the non-existence
of the predicted \cite{5,6,9,11} instability or low frequency mode in the
vicinity of the first roton minimum. We note that the reported Raman
scattering experiments \cite{23} are inconclusive because of the low $r_{s}$
and relatively high $k$ values involved. Recent advances in fabricating high 
$r_s$ samples \cite{24} and small layer separation should render the
suggested experiments feasible.

In summary, we have obtained a comprehensive picture of the collective mode
structure of an electronic bilayer in the strongly coupled liquid phase.
This structure is qualitatively different both from that of the weakly
coupled bilayer electron gas (describable in the RPA) and from the phonon
spectrum of the bilayer solid and exhibit a number of remarkable features
which should be experimentally verifiable. The calculation avoids the
pitfalls of earlier approaches which stem from the inconsistencies in the
approximations used for the interlayer and for the intralayer PCF-s, and
from the violation of the $3^{rd}$ frequency-moment-sum rule.

We wish to thank Dan Dubin for making his unpublished results available to
us. GK is grateful to Tom O'Neil for making possible his stay at UCSD, where
this work was started. We thank Hongbo Zhao for help in preparation of the
Figures. This work has been partially supported by DOE Grants
DE-FG02-98ER54501 and DE-FG02-98ER54491.

\begin{figure}[tbp]
\caption{ The four principal modes for $\Gamma=40$ ($r_s=20$): (a) $d/a=0.3$
(b) $d/a=1.0$. The shaded region is the pair excitation continuum. For $%
d/a=1.0$ ($\omega_{min} \sim 0.45 \omega_0$) the diffusional domain in which
the shear mode is non-propagating and the out-of-phase plasmon exhibits an
RPA-like behavior is indicated. }
\end{figure}

\begin{figure}[tbp]
\caption{ The gapped out-of-phase ("acoustic") plasmon for different layer
separations. The shaded region is the pair excitation continuum. In the $%
\omega<\omega_{min}$ domain the $d/a=1.0, 1.5, 2.0$ gap values are spurious
and the dispersion should resemble the RPA acoustic type behavior. }
\end{figure}

\begin{figure}[tbp]
\caption{ Gap ($\omega(0)$) value for $\Gamma=40$ and $\Gamma=30$ as
functions of the layer separation. Below the estimeted $\omega_{min}(\Gamma)$
value the gap is spurious. The inset compares $\omega(0)$ and $%
\omega(\infty) $ for $\Gamma=40$. }
\end{figure}


\begin{references}
\bibitem{1}  Swierkowski, L., D. Neilson and J. Szymanski, Phys. Rev. Lett. 
{\bf 67}, 240 (1991).

\bibitem{2}  Rapisada, F. and G. Senatore, {\it Strongly Coupled Coulomb
Systems}, edited by G. Kalman, K. Blagoev and M. Rommel, Plenum Press (in
press).

\bibitem{3}  Valtchinov, V., G. Kalman and K.B. Blagoev, Phys. Rev. {\bf E 56%
}, 4351 (1997).

\bibitem{4}  Goldoni, G. and F.M. Peeters, Phys. Rev. {\bf B 53}, 4591
(1996).

\bibitem{5}  (a). Swierkowski, L., D. Neilson and J. Szymanski, Austr. J.
Phys., {\bf 46}, 423 (1992); (b) Neilson, D., L. Swierkowski, J. Szymanski
and L. Liu, Phys. Rev. Lett. {\bf 71}, 4035 (1993), {\it erratum}, Phys.
Rev. Lett. {\bf 72}, 2669 (1994); (c) Liu. L., L. Swierkowski, D. Neilson
and J. Szymanski, Phys. Rev. {\bf B 53}, 7923 (1996).

\bibitem{6}  Gold, A., Z. Phys. {\bf B 86}, 193 (1992): {\bf 90}, 173
(1993): Phys. Rev. {\bf B 47}, 6762 (1993): Gold, A. and L. Calmels, {\it %
ibid}, {\bf 48}, 11 622 (1993): Gold, A., Z. Phys. {\bf B 95}, 341 (1994);
97, 119 (1995).

\bibitem{7}  Zhang, C. and N. Tzoar, Phys. Rev. {\bf A 38}, 5786 (1988);
Zhang, C., Phys. Rev. {\bf B 49}, 2939 (1994).

\bibitem{8}  Golden, K.I. and G. Kalman, {\it physica status solidi (b)} 
{\bf 180}, 533 (1993); Kalman, G., Y. Ren and K.I. Golden, {\it %
Contributions to Plasma Physics} {\bf 33}, 449 (1993); Kalman, G. and K.I.
Golden, in {\it Condensed Matter Theories} {\bf 8}, edited by L. Blum and
B.S. Malik (Plenum Press) (1993); Golden, K.I. in {\it Modern Perspectives
in Many-Body Theory}, edited by M.P. Das and J. Mahanty (World Scientific
Press) (1994); Kalman, G., Y. Ren and K.I. Golden, Phys. Rev. Rapid Comm. 
{\bf B 50}, 2031 (1994).

\bibitem{9}  Lu, Dexin. K.I. Golden, G. Kalman, Ph. Wyns, L. Miao and X.L.
Shi, Phys. Rev. {\bf B 54}, 11457 (1996).

\bibitem{10}  Golden, K.I., G. Kalman., L. Miao and R.R. Snapp, Phys. Rev. 
{\bf B 55}, 16349 (1997).

\bibitem{11}  Moudgil, R.K., P.K. Ahluvalia and K.N. Pathak, Phys. Rev. {\bf %
B56}, 14776 (1997).

\bibitem{12}  Dubin, D.H.E., Phys. Rev. Lett. {\bf 66}, 2076 (1991); {\bf 71}%
, 2753 (1993); Phys. Fluids {\bf B 5}, 295 (1993); (b) Mitchell, T.B.,
Bollinger J.J, D.H.E. Dubin, X.-P. Huang, W.M. Itano and R.H. Bundham,
''Direct observation of structural phase transition in planar crystallized
ion plasmas'' (unpublished).

\bibitem{13}  Kalman, G. and K.I. Golden, Phys. Rev. {\bf B 57} , 8834
(1998).

\bibitem{14}  Iwamoto, N., E. Krotschek and D. Pines, Phys. Rev. {\bf B 29},
3936 (1984); Iwamoto, N. Phys. Rev. {\bf A 30}, 3289 (1984).

\bibitem{15}  Kalman, G. and K.I. Golden, Phys. Rev. {\bf A 41}, 5516 (1990).

\bibitem{16}  Golden, K.I. , G. Kalman and Ph. Wyns, Phys. Rev. {\bf A 41},
6940 (1990).

\bibitem{17}  (a) Golden, K.I. , G. Kalman and Ph. Wyns, Phys. Rev. {\bf A 46%
}, 3454 (1992; (b) 3463 (1992).

\bibitem{18}  Fetter, A.L., Ann. Phys. (N.Y.) {\bf 88}, 1 (1974).

\bibitem{19}  Das Sarma, S. and A. Madhukar, Phys. Rev. {\bf B 23},
805(1981).

\bibitem{20}  Hori, J., {\it Spectral Properties of Disordered Lattices},
Pergamon Press, NY, (1967).

\bibitem{21}  Totsuji, H. and H. Kakeya, Phys. Rev. {\bf A 22}, 1220 (1980).

\bibitem{22}  Hansen, J.P., I.R. McDonald and E.L. Pollock, Phys. Rev. {\bf %
A 11}, 1025 (1975).

\bibitem{23}  Olego, D., A. Pinczuk, A.C. Gossard and W. Wiegmann, Phys.
Rev. {\bf B 25}, 7867 (1982); Fasol, G., N. Mestres, H.P. Hughes, A. Fischer
and K. Ploog, Phys. Rev. Lett. {\bf 56}, 2517 (1986).

\bibitem{24}  Shapira, S., U. Sivan, P.M. Solomon, E. Buchstab, M. Tischler
and G.B. Yoseph, Phys. Rev. Lett. {\bf 77}, 3181 (1996); S. Shapira {\it et
al} in {\it Strongly Coupled Coulomb Systems}, edited by G. Kalman, K.
Blagoev and M. Rommel, Plenum Press (in press).
\end{references}
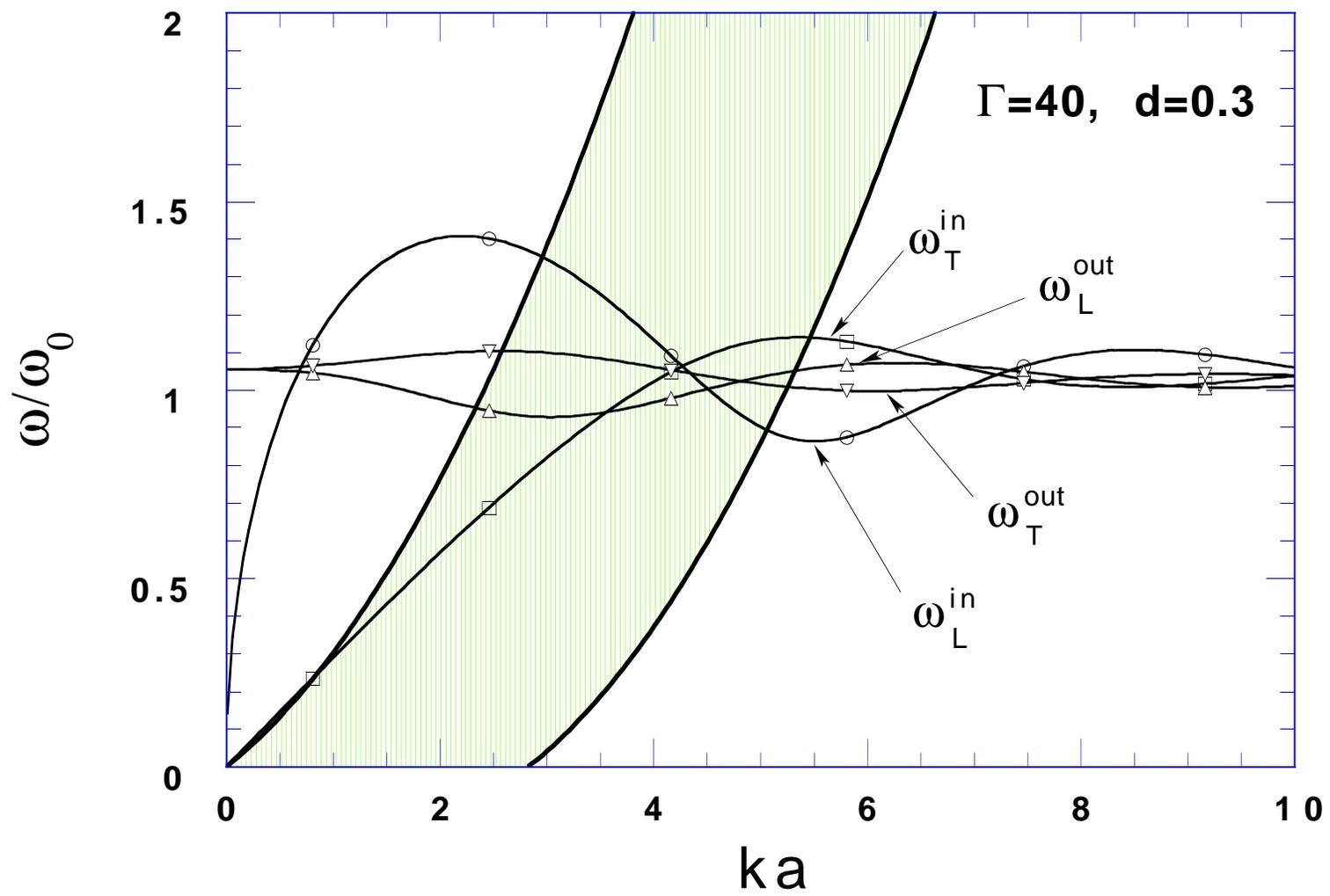
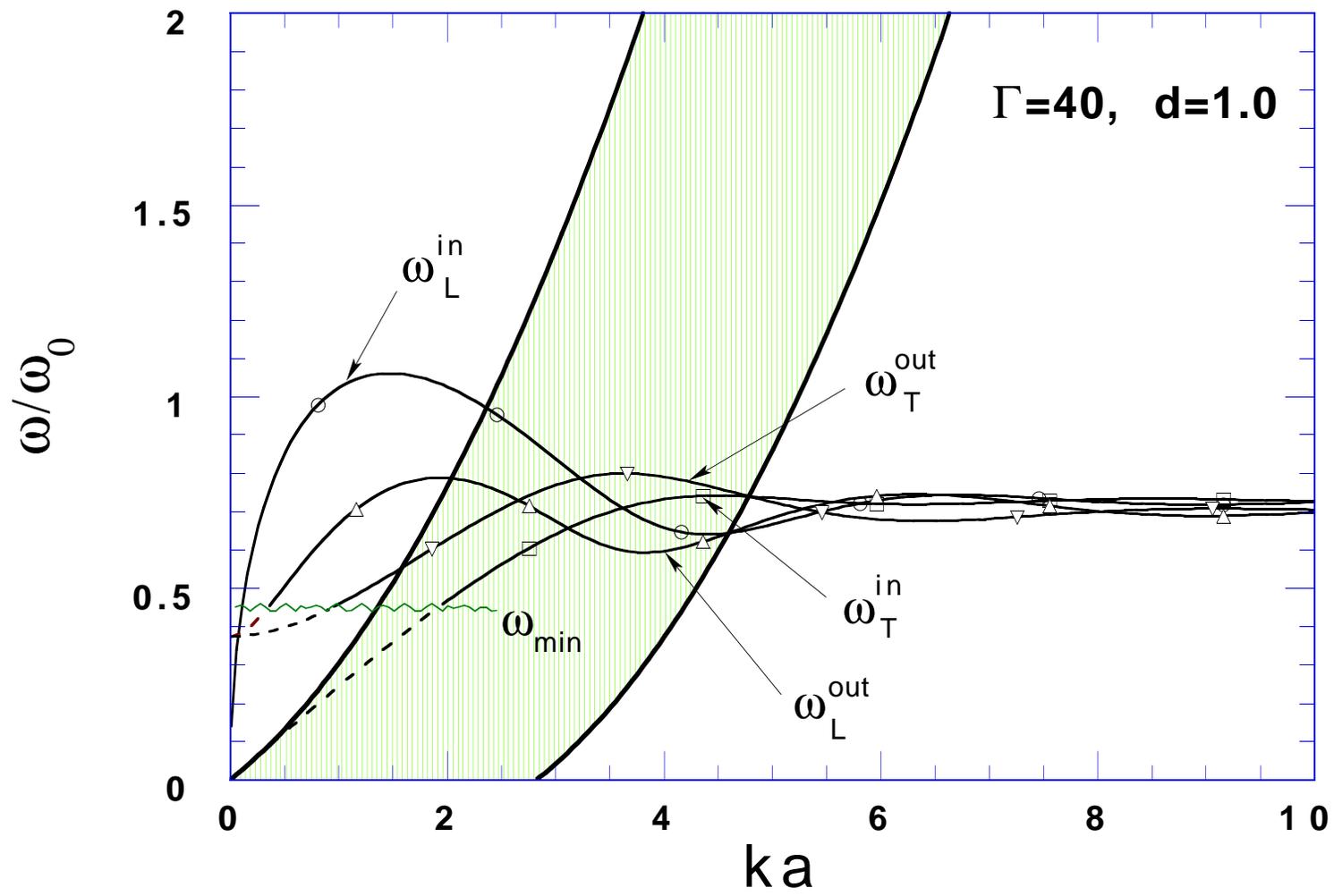
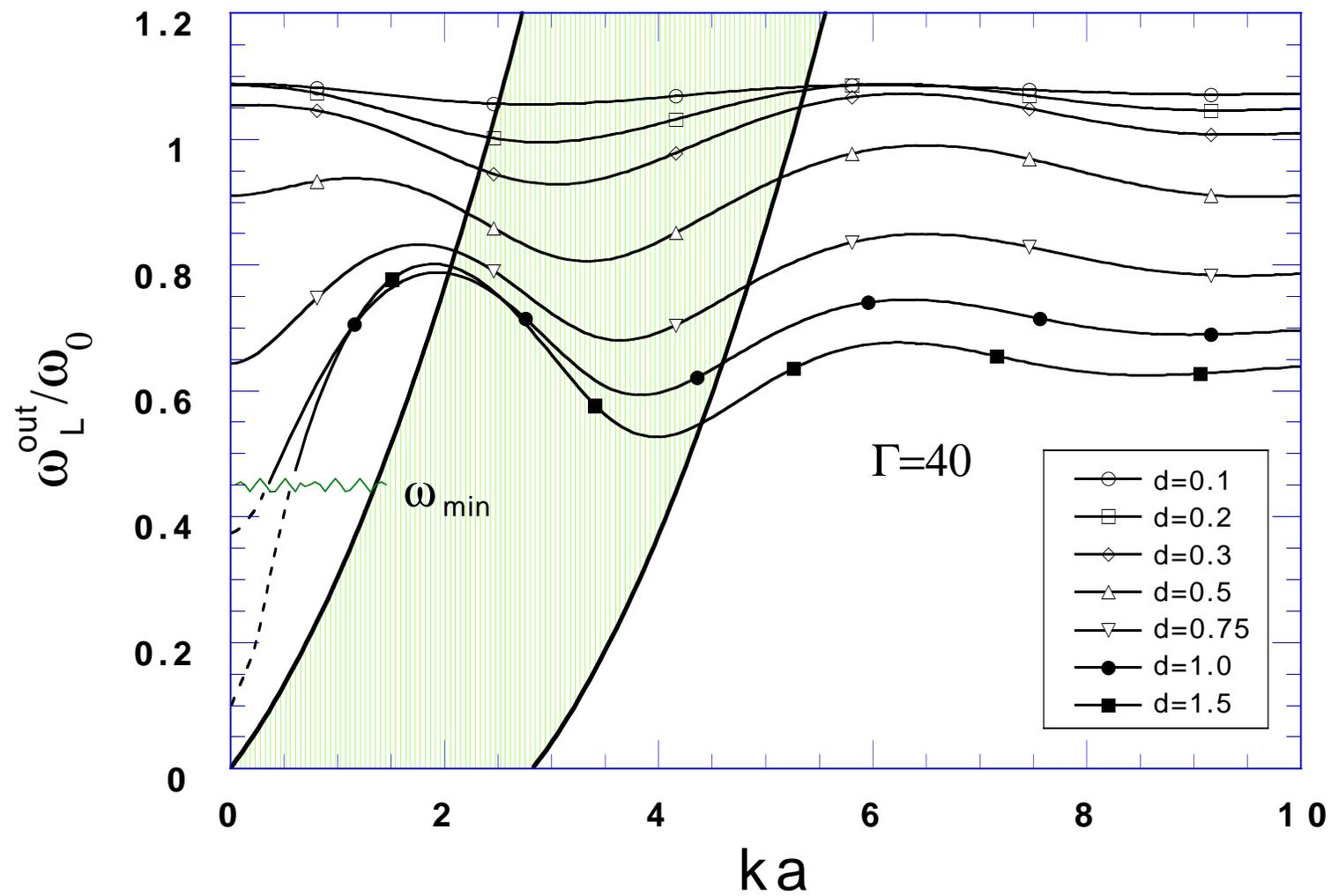
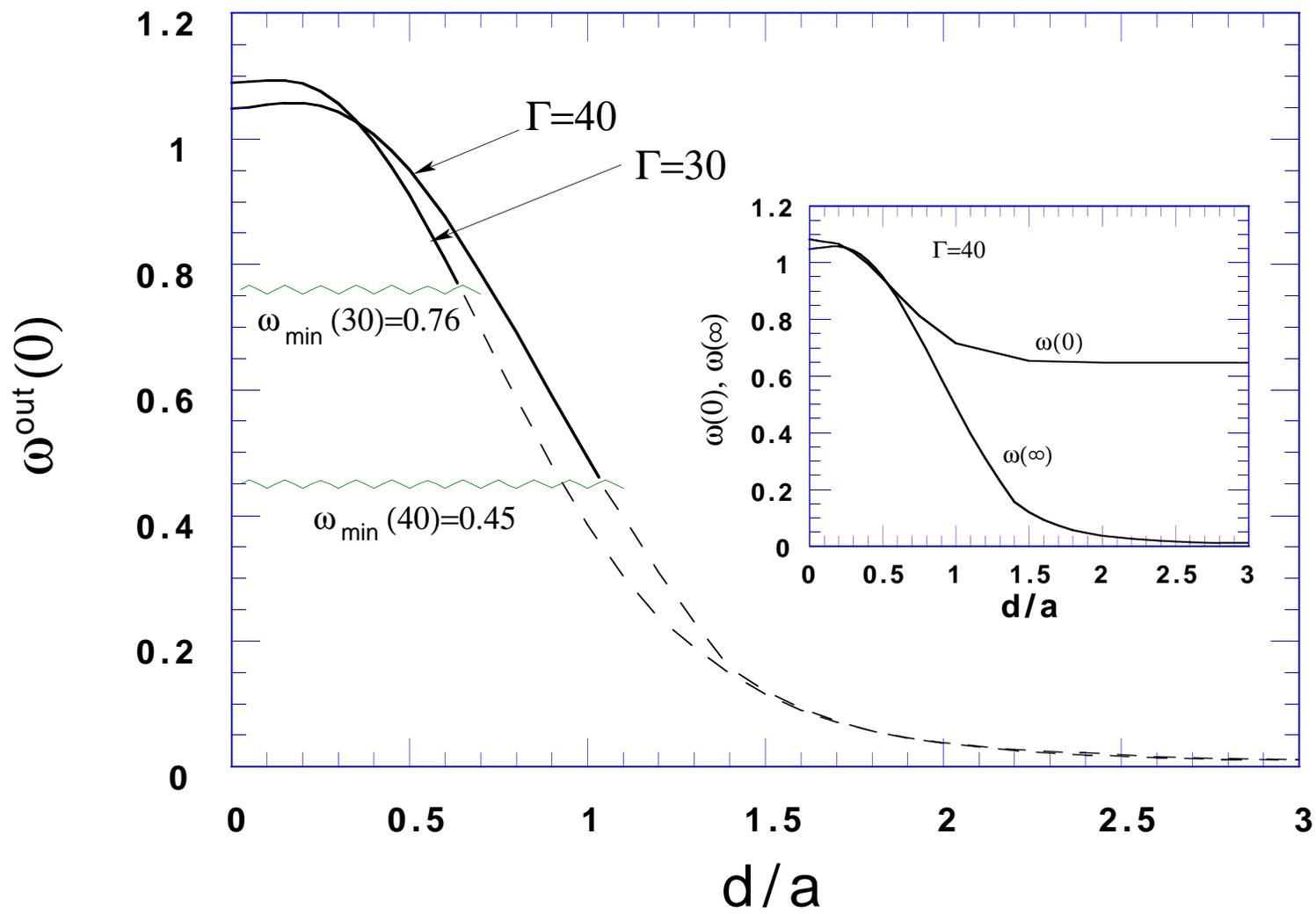
\end{document}